\documentclass[prd,preprint,tightenlines,floatfix,preprintnumbers,nofootinbib,eqsecnum]{revtex4}
 \usepackage[dvips,final]{graphicx}
  \usepackage{amssymb}
   \usepackage{amsmath}
    \usepackage{epsfig}
     \usepackage{bm}
      \usepackage{pifont}

\newcommand{\be}{\begin{equation}}\newcommand{\ee}{\end{equation}}       
\newcommand{\bd}{\begin{displaymath}}\newcommand{\ed}{\end{displaymath}} 
\newcommand{\bit}{\begin{itemize}}\newcommand{\eit}{\end{itemize}}       

\newcommand{\ben}{\begin{enumerate}}\newcommand{\een}{\end{enumerate}}   
\newcommand{\baa}{\begin{array}{lll}}\newcommand{\eaa}{\end{array}}      
\newcommand{\ba}{\begin{eqnarray}}\newcommand{\ea}{\end{eqnarray}}       
\def\kp{\relax\ifmmode{k_{\perp}}\else{$k_{\perp}${ }}\fi}               
\newcommand{\itxt}[1]{\relax\ifmmode{\text{#1}}\else{#1{}}\fi}           
\newcommand{\gev}[1]{\relax\ifmmode{\text{GeV}^{#1}}                     
                     \else{GeV$^{#1}${ }}\fi}                            
\def\Gev{\relax\ifmmode{\text{GeV}}\else{GeV{ }}\fi}                     
\def\Mev{\relax\ifmmode{\text{MeV}}\else{MeV{ }}\fi}                     
\newcommand{\footn}{\footnotesize}                                       
\newcommand{\la}{\label}                                                 
\newcommand{\Ds}{\displaystyle}                                          
\newcommand{\va}[1]{\langle{#1}\rangle}                                  
\newcommand{\nn}{\nonumber}                                              
\newcommand{\xx}{\left(\bar x\rightarrow x\right)}                       
\newcommand{\xxyy}                                                       
{\left(\bar x\rightarrow x,\,\bar y\rightarrow y\right)}                 
\def\MSbar{\relax\ifmmode\overline                                       
          {\textrm{MS}}\else{$\overline{\textrm{MS}}${}}\fi}             
\def\as{\relax\ifmmode \alpha_s\else{$\alpha_s${}}\fi}                   
\def\abar{\relax\ifmmode{\bar{a}}\else{$\bar{a}${}}\fi}                  
\newcounter{myfig}                                        
\newcommand{\myfig}{\refstepcounter{myfig}}               

\begin{document}
\preprint{\vbox{\hbox{JINR-E2-2001-60}\hbox{RUB-TPII-03/01}}}
\title{QCD-based pion distribution amplitudes confronting experimental
       data}
\author{{\bf A.~P.~Bakulev},${}^{1}$\thanks{Email:
        bakulev@thsun1.jinr.ru}
        {\bf S.~V.~Mikhailov},${}^{1}$\thanks{Email:
        mikhs@thsun1.jinr.ru}
        {\bf N.~G.~Stefanis}${}^{2}$\thanks{Email:
        stefanis@tp2.ruhr-uni-bochum.de}
        }
\address{${}^{1}$ Bogoliubov Laboratory of Theoretical Physics,  \\
                  Joint Institute for Nuclear Research,          \\
         \vspace{5mm}                  
                  141980, Moscow Region, Dubna, Russia           \\
         ${}^{2}$ Institut f\"ur Theoretische Physik II          \\
                  Ruhr-Universit\"at Bochum                      \\
                  D-44780 Bochum, Germany                        \\
         }

\begin{abstract}
We use QCD sum rules with nonlocal condensates to re-calculate
more accurately the moments and their confidence intervals of the
twist-2 pion distribution amplitude including radiative
corrections. We are thus able to construct an admissible set of
pion distribution amplitudes which define a reliability region in
the $a_{2}$, $a_{4}$ plane of the Gegenbauer polynomial expansion
coefficients. We emphasize that models like that of Chernyak and
Zhitnitsky, as well as the asymptotic solution, are excluded from
this set. We show that the determined $a_{2}$, $a_{4}$ region
strongly overlaps with that extracted from the CLEO data by
Schmedding and Yakovlev and that this region is also not far from
the results of the first direct measurement of the pion valence
quark momentum distribution by the Fermilab E791 collaboration.
Comparisons with recent lattice calculations and instanton-based
models are briefly discussed.
\end{abstract}
\pacs{11.10.Hi,12.38.Bx,12.38.Lg,13.40.Gp}
\maketitle

\section{Introduction}
The pion distribution amplitude (DA) of twist-2,
$\varphi_\pi(x,\mu^2)$, i.e., the integral over transverse momenta
of the renormalized valence-quark wave function on the light cone,
is a gauge- and process-independent characteristic of the pion
and, due to factorization theorems \cite{CZ77,ERBL79}, it enters
as the central input various QCD calculations of hard exclusive
processes. A reliable derivation of the pion DA from first
principles in QCD is therefore an outstanding problem of paramount
importance, given that it contains all of the bound-state dynamics
and specifies in a universal way the longitudinal momentum $xP$
distribution of the valence quarks in the pion with momentum $P$
(see, e.g., \cite{CZ84} for a review),
\be \va{0\mid \bar
d(z)\gamma^{\mu}\gamma_5 u(0)\mid \pi(P)}\Big|_{z^2=0}
 = i f_{\pi}P^{\mu}
    \int^1_0 dx e^{ix(zP)}\ \varphi_{\pi}(x,\mu^2)\ .
\ee 

Recently, the CLEO collaboration \cite{CLEO98} has measured the
$\gamma^{*}\gamma \to \pi^{0}$ form factor with high precision.
These data sets have been processed by Schmedding and Yakovlev
(S\&Y) \cite{SchmYa99} using light-cone QCD sum rules and
including perturbative QCD contributions in the NLO approximation
(NLA), to obtain useful constraints on the shape of the pion DA in
terms of confidence regions for the Gegenbauer coefficients
$a_{2}$ and $a_{4}$. On the other hand, the first direct
measurement of the transverse momentum distribution in the pion
via diffractive dissociation into di-jets by the Fermilab E791
Collaboration \cite{Ash00}, supplemented by new lattice
results~\cite{Dal01} for the same quantity, also provides the
possibility to deduce the shape of the pion DA (modulo inherent
method uncertainties).

These important findings inevitably raise the question of whether
pion DAs, derived from first principles of QCD, can confront them
successfully. The present analysis is a targeted investigation of
these issues, the focus being on pion DAs, reconstructed from QCD
sum rules with nonlocal condensates (NLC-SR).

This approach, developed in \cite{MR86,MR89,BR91,MS93} by
A.~Radyushkin and two of us (A.B. and S.M.) provides a reliable
method to construct hadron DAs that inherently accounts for the
fact that quarks and gluons can flow through the QCD vacuum with
{\it non-zero momentum} $k_q$. This means, in particular, that the
{\it average} virtuality of vacuum quarks
$\langle k_{q}^{2} \rangle = \lambda_{q}^{2}$
is not zero, like in the local sum-rule approach \cite{CZ84}, but
sizeable. The (vacuum) non-locality parameter is the only one
involved in the NLC-SR method, and it has been estimated within the
QCD sum-rule approach from the mixed quark-gluon condensate of
dimension 5:
\be
\label{lambda}
 \lambda_{q}^{2}= \frac{\va{\bar{q}(0)\nabla^{2} q(0)}}
                   {\va{\bar{q}(0)q(0)}}
            \begin{array}{c}
             \text{\footn in chiral}\\
             =\\
             \text{\footn limit}
             \end{array}
              \frac{\va{\bar{q}(0)
                        \left(i g\,\sigma_{\mu \nu}G^{\mu \nu}
                        \right)q(0)}}
                   {2\va{\bar{q}(0)q(0)}}
   = \left\{\begin{array}{ll}
         0.4 \pm 0.1~\gev{2}& \text{\cite{BI82}}\\
         0.5 \pm 0.05~\gev{2}&\text{\cite{Piv91}}
             \end{array}
     \right.\ .
\ee 

Lacking an exact knowledge of NLC of higher dimensionality, the
non-locality can only be taken into account in the form of an ansatz.

A simple model is provided by a Gaussian-like behavior of NLC
\cite{MR89,BR91}, whereas $\lambda_{q}^{-1}$ reveals itself as the
typical quark-gluon correlation length in the QCD vacuum. Hence,
we set in coordinate space,
$
 M_{S}(z^{2})\equiv \langle \bar q(0)
 E(0,z)q(z)\rangle$ $\Ds \sim \langle \bar q q \rangle
 \exp\left(- \lambda_{q}^{2} |z^{2}|/8\right)
$,\footnote{%
\footnotesize Here, as usual, $E(0,z)=P\exp(i \int_0^z dt_{\mu}
A^a_{\mu}(t)\tau_a)$ is the Schwinger phase factor required for
gauge invariance.} while the coordinate behavior of other NLCs
looks more complicated.

Estimates of $\lambda_{q}^{2}$ from instanton approaches
\cite{PW96,DEM97} are somewhat larger:
$\lambda_{q}^{2} \approx 2/\rho_{c}^{2} \geq 0.6~\gev{2}$
(where $\rho_{c}$ is the characteristic size
of the instanton fluctuation in the QCD vacuum:
$0.33 \leq \rho_{c} \leq 0.6$~Fm),
while lattice calculations \cite{DDM99,Meg99,DEJM2000} confirm
qualitatively the Gaussian law for the $M_S(z^2)$ decay, yielding
values close to those given in Eq.(\ref{lambda}). More calculational
details are relegated to \cite{MR89,BM95,BM98,BM00}.

The main results of the present analysis can be summarized as
follows. The moments of the pion DA have been re-calculated more
accurately and, more importantly, their confidence levels have
been determined. In this way, we are able to construct a whole
``bunch'' (a spectrum) of pion DAs, allowed by the NLC-SR for
different values of $\lambda_{q}^{2}$, as noticed above. The
important thing to be emphasized is that the range of this
spectrum in the $a_{2}$, $a_{4}$ plane strongly overlaps with the
$95\%$ and also with the $68\%$ region of Schmedding and Yakovlev
\cite{SchmYa99}. Perhaps somewhat surprisingly, the shape of the
optimum pion DA, we have determined, is not ``dromedary''-like, as
the asymptotic solution, but has two maxima (i.e., it is
``camel''-like). Nevertheless, we stress, it is not
Chernyak--Zhitnitsky (CZ) like either because in the endpoint
region it is strongly suppressed.

\section{Spectrum of pion DA\lowercase{s} from NLC-SR}
\textbf{Revision of pion DA moments.}
The NLC-SR for the DAs of the pion and the effective
$A_{1}+\pi^{\prime}$-resonance, $\varphi_\pi(x)$ and
$\varphi_{A_1}(x)$, respectively, which appear in the ``axial''
channel, were constructed in~\cite{MR86,MR89} and were analyzed in
\cite{MR89,BM98}.
Here we display only the sum rule itself:
\ba&\Ds \label{eq:SRDA}
 \left(f_{\pi}\right)^2\varphi_\pi(x)
 + \left(f_{A_1}\right)^2\varphi_{A_1}(x) e^{-m^2_{A_1}/M^2}
 = \int_{0}^{s_0^A}\rho^{pert}(x;s)e^{-s/M^2}ds
 + \Delta\Phi_S(x;M^2)
 \phantom{.}& \\&\Ds\phantom{.}
 + \Delta\Phi_V(x;M^2) + \Delta\Phi_{T_1}(x;M^2)
 + \Delta\Phi_{T_2}(x;M^2) + \Delta\Phi_{T_3}(x;M^2)
 + \Delta\Phi_G(x;M^2)&
\nonumber \ea
and re-estimate the moments and their error bars.
The NLC contributions encoded in $\Delta\Phi_{\Gamma}(x;M^2)$ on
the rhs of the SR, as well as the explanation of the individual
terms are discussed in \cite{MR89,BM98}. The corrected form of the
contribution for $\Delta\Phi_{T_1}(x;M^2)$ (originating from the
NLC $\va{\bar{q}Gq}$) and also the main contribution to the rhs of
the SR, which is accumulated in the $\Delta\Phi_{S}(x;M^2)$ term
(owing to the NLC $M_S$), are shown in Appendix~A.

\begin{figure}[thb]
 \centerline{\includegraphics[width=\textwidth]{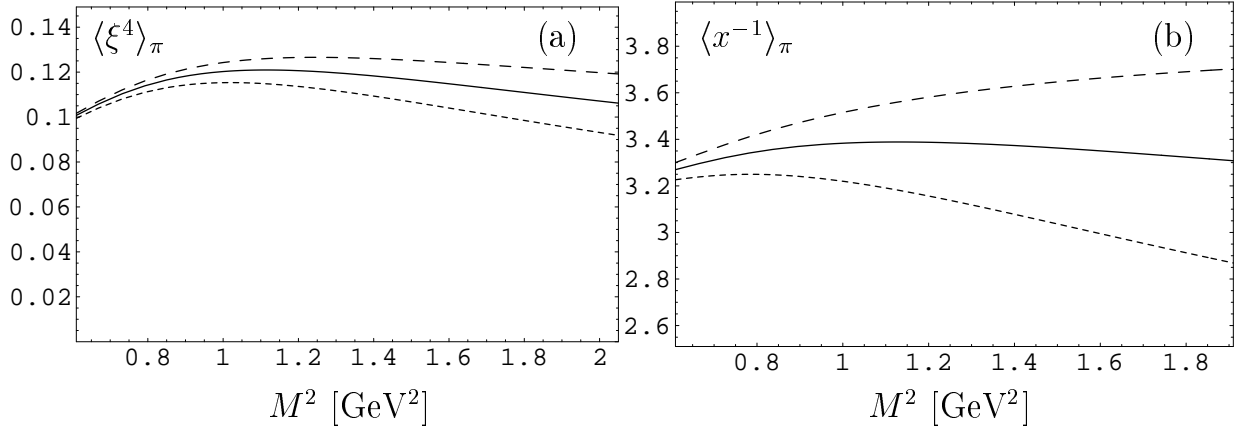}}
   \vspace*{1mm}\myfig{\label{fig:moments}}
    \caption{\footnotesize
   (a) Original SR for the moment $\langle \xi^4\rangle_{\pi}$ and
   (b) ``daughter SR'' for the inverse moment $\langle
   x^{-1}\rangle_{\pi}^\itxt{SR}$, both as functions of the Borel
   parameter $M^2$, as obtained from the NLC-SRs.
   Both kinds of SRs were processed by including the $A_1$-meson with
   $s_0=2.2~\gev{2}$.
   For the depicted range of $M^2$, the fidelity and stability windows
   almost coincide.
   Solid lines correspond to the optimal threshold $s_0$, whereas
   broken lines denote the results for the upper (long-dashed line)
   and lower (short-dashed line) moment limit with a $10\%$-variation
   of $s_0$.}
 \end{figure}

Both the sensitivity and the stability of the NLC-SR are considerably
improved relative to the standard SR (but nevertheless we still
assume a rather conservative size of the errors of the order of
$10\%$).
This allows us to determine the first ten moments
$\langle\xi^N\rangle_\pi \equiv \int_{0}^{1}
\varphi_\pi(x)(2x-1)^N dx$ of the pion DA quite accurately. This
is illustrated in Fig.~\ref{fig:moments}(a), where the curves for
$\langle \xi^4 \rangle_{\pi}$ are represented as functions of the
Borel parameter $M^2$ plotted within the range of the stability
window of the SR. The solid line (optimum moment) is determined
from the NLC-SR (\ref{eq:SRDA}), for $s_{0}=2.2$~GeV${}^{2}$,
while the broken lines give the results for thresholds varied by
$10\%$ around this value, which turns out to be practically
independent of the moment order $N$ \cite{MR89,BM98}. Then, the
fidelity windows are: $0.5~\gev{2} \le M^2 \le 2.0~\gev{2}$ for
all $N=0, 2, \ldots, 10$. Notice that the range of stability and
the Borel fidelity windows almost coincide with each other,
starting for all $N$ at $M^2\approx 0.6~\gev{2}$.
\input{psbox}
\vspace*{1mm}
\vspace*{-10mm}

\vspace*{5mm}
\noindent\hspace*{0.025\textwidth}
\begin{minipage}{0.95\textwidth}
\footnotesize
  \begin{tabular}{|cl||c|c|c|c|c|c|c|}\hline
 & & & & & & & &\\
\multicolumn{2}{|c||}{Type of SR}
   & $f_{M}\left(\text{GeV}\right)$
     &$\hspace{0.1mm}N=2\hspace{0.1mm}$
       &$\hspace{0.1mm}N=4\hspace{0.1mm}$
         &$\hspace{0.1mm}N=6\hspace{0.1mm}$
           &$\hspace{0.1mm}N=8\hspace{0.1mm}$
             &$\hspace{0.1mm}N=10\hspace{0.1mm}$
               &$\hspace{0.1mm}\va{x^{-1}}\hspace{0.1mm}$ \\
 & & & & & & & &\\ \hline \hline
\multicolumn{2}{|c||}{\strut\vphantom{\vbox to 6mm{}}
  Asympt. DA$_{\vphantom{\vbox to 4mm{}}}$}
    &    & $0.2$  & $0.086$
        & $0.047$ & $0.030$ & $0.021$ & $3$
   \\  \hline \hline
 {\strut\vphantom{\vbox to 6mm{}} NLC SR ~\cite{MR89}
$_{\vphantom{\vbox to 4mm{}}}$:}

 &$\pi $
   &$0.131$&$0.25$&$0.12$
         &$0.07$& -- & -- & --
  \\ \cline{3-9}
 {\strut\vphantom{\vbox to 6mm{}} NLC SR ~\cite{BM98}
  $_{\vphantom{\vbox to 4mm{}}}$:}
 & $\pi $
   & $0.131(2)$ & $0.25(1)$ & $0.110(7)$
         & $0.054(3)$ & $0.031(2)$ & 0.0217(7) & $2.75(5)^\itxt{SR}$
      \\ \hline \hline
 {\strut\vphantom{\vbox to 6mm{}} NLC this work
  $_{\vphantom{\vbox to 4mm{}}}$:}
 &$\pi $
   & $0.131(8)$
                & $0.265(20)$
                            & $0.115(12)$
         & $0.061(8)$
                      & $0.037(5)$
                                   & $0.024(4)$
                                               & $3.35(32)^\itxt{SR}$
    \\ \cline{3-9}
 {\strut\vphantom{\vbox to 6mm{}} NLC this work
  $_{\vphantom{\vbox to 4mm{}}}$:}
 & $A_1+\pi^{\prime}$ & $0.210(17)$ & $0.21(2)$ & $0.116(12)$

                & $0.078(8)$ & $0.055(6)$ & $0.042(5)$ & $3.6(4)^\itxt{SR}$
    \\ \cline{3-9}
 {\strut\vphantom{\vbox to 6mm{}} CZ SR~\cite{CZ82,CZ84}
  $_{\vphantom{\vbox to 4mm{}}}$:}
 &$\pi $  &$0.131$   &  $0.40$   &$0.24$   & --   & --  & --   & --
    \\ \hline \hline
\end{tabular}
\vspace*{3mm}

\noindent\textbf{Table 1}. The moments $\langle \xi^N
\rangle_{M}(\mu^2)$ determined at $\mu^2 \sim 1~\gev{2}$ with
associated errors put in parentheses. Recall that CZ give all
moments normalized at the lower scale $\mu_0^2=0.5~\gev{2}$.
\end{minipage}
\vspace*{3mm}

The important convolution $\Ds \langle x^{-1}\rangle_{\pi}=
\int^1_0 \frac{\varphi_\pi(x)}{x}dx$, appears in the perturbative
calculation of the $\gamma^{*}\gamma \to \pi^0$ process. We
construct a ``daughter SR'' directly for this quantity from
Eq.~(\ref{eq:SRDA}) by integrating its rhs with the weight $1/x$.
Due to the smooth behavior of the NLC at the end points $x=0,1$,
this integral is well defined, supplying us with an independent
SR, with a rather good stability behavior of $\langle
x^{-1}\rangle_{\pi}^\itxt{SR}(M^2)$, as one sees from
Fig.~\ref{fig:moments}(b). To understand this behavior, let us
recall an important feature of the NLC, encoded in
$\Delta\Phi_{\Gamma}(x;M^2)$. This contribution is suppressed in
the vicinity of the end-points $x=0,1$, the range of suppression
being controlled by the value of the non-locality parameter
$\lambda_q^2$. The larger this parameter, at fixed resolution
scale $M^{2} > \lambda_q^2$, the stronger the suppression of the NLC
contribution. Similarly, an excess of the value of
$\langle x^{-1}\rangle_{\pi}$ over 3 (asymptotic DA) is also
controlled by the value of $\lambda_q^2$, becoming smaller with
increasing $\lambda_q^2$. Note, that instanton-based models
demonstrate the same behavior of the DA with a non-locality
parameter proportional to $ \rho_c^{-2}$~\cite{ADL00}.

From Table 1 one infers that the mean values of the re-calculated
moments are close to the old ones, whereas the value of $\langle
x^{-1}\rangle_{\pi}^\itxt{SR}$ changes significantly. The reason
is the corrected term $\Delta\Phi_{T_1}(x;M^2)$ (in the NLC-SR),
which allows us now to reconcile this quantity with the value
obtained by calculating it with the model DAs (see below). The new
estimates of the error bars in Table 1 have been formed from
different sources. Note that the most significant error (about
$50\%$) stems from the uncertainties of the parameters of the
$A_1+\pi^{\prime}$-resonance. \vspace*{3mm}

\begin{figure}[thb]
 \centerline{\includegraphics[width=\textwidth]{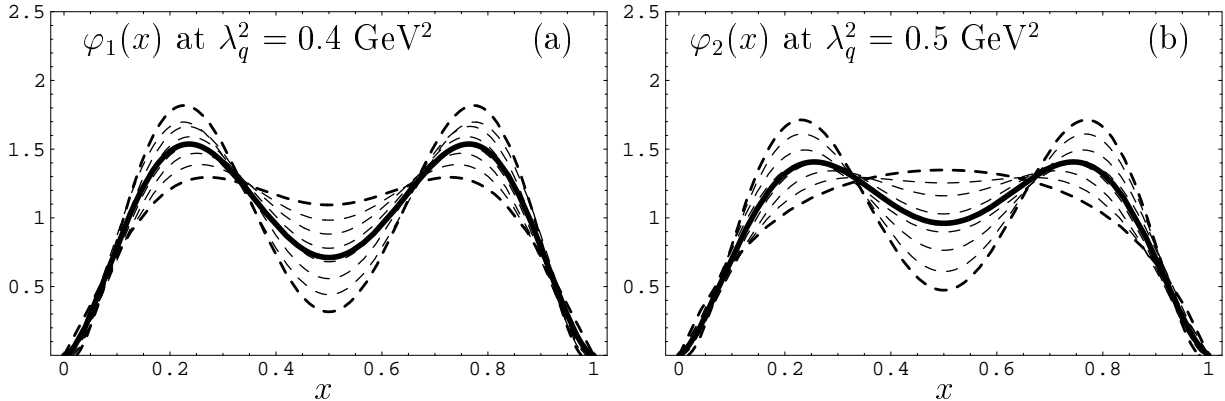}}
   \vspace*{1mm}\myfig{\label{fig:DArange}}
     \caption{\footnotesize
   Graphical representation of $\varphi_1(x)$ (part (a)) and
   $\varphi_2(x)$ (part (b)).
   The thick solid lines in both plots denote
   $\varphi^{\text{opt}}_{1,2}(x)$, i.e., the best fit to the
   determined values of the moments (see Table 1), whereas dashed
   lines illustrate admissible options with $\chi^2\ \leq 1$.}
 \end{figure}

\textbf{Spectrum of admissible pion DAs.} Models for the pion DA,
in correspondence to the moments in Table 1, can be constructed in
different ways \cite{MR89,BM98}. However, it appears that
two-parameter models, the parameters being the Gegenbauer
coefficients $a_{2}$ and $a_{4}$ (as also used in \cite{SchmYa99}),
enable one to fit all the moment constraints for
$\langle \xi^{N} \rangle_\pi$ given in the Table, as well as to
reproduce the value of $\va{x^{-1}}_\pi$ within the quoted error
range.\footnote{\footnotesize{Note that SY {\it assume} that
$a_{n>4}$ are small, while in our approach these coefficients have
been {\it calculated} up to $N=10$ and found to be numerically
negligible.}} The optimum model DA is \ba \la{optG}
 \varphi_1^{\text{opt}}(x) &=& \varphi^{\rm as}(x)
 \left[1+a_2^{\itxt{opt1}} \cdot C^{3/2}_2(2x-1)
        +a_4^{\itxt{opt1}} \cdot C^{3/2}_4(2x-1) \right]\ ,\\
 a_2^{\itxt{opt1}}&=& + 0.188\ ,\quad
 a_4^{\itxt{opt1}}\ =\ - 0.130\ ,\quad
 \text{with}~~\chi^2 \approx 10^{-3}\ ,
 \la{opt_a24}
 \ea 
yielding $\va{x^{-1}}_1^{\text{opt}} = 3.174$. But one can
construct a whole admissible set (a spectrum),
$\left\{\varphi_1(x;a_2,a_4)\right\}$, of such models by demanding
that the associated moments lie inside the error bars, presented
in Table 1. This set of DAs approximately corresponds to those
parameters $(a_2,a_4)$, which satisfy $\chi^2 \leq 1$. In this
way, we obtain a ``bunch'' of admissible DA profiles, shown in
Fig.~\ref{fig:DArange}(a) in terms of dashed lines, in addition to
the optimum one (thick solid line). For these DAs, the
corresponding values of $\va{x^{-1}}_{\varphi_1}$ vary in the
interval \be \label{invers1} 3.08 \leq \va{x^{-1}}_{\varphi_1}
\leq 3.24
 ~~~~\text{vs}~~~~
   3.03 \leq \va{x^{-1}}^{SR}_{\pi 1} \leq 3.67\ .
\ee 
Let us mention at this point that in the standard approach the
value of $\va{x^{-1}}(\mu^2)$ cannot be determined from the
\textit{sum rule itself} due to the presence of singular
$\delta(x)$-terms. On the other hand, for the CZ {\it model} DA,
one finds $\va{x^{-1}}_{\rm CZ} (\mu^2) \approx 5.1$ at
normalization scale $\mu^2 \approx (0.7)^2~\gev{2}$ or $\approx
4.44$ at the scale $\mu^2 \approx 1~\gev{2}$.
Meanwhile, our estimates of the inverse moment in (\ref{invers1}),
as well as the value of $a_2$ in (\ref{opt_a24}) are in a good 
agreement with the estimates 
$\va{x^{-1}} ( 1~\gev{2}) = 3.3\pm 0.3$, and 
$a_2( 1~\gev{2})=0.1 \pm 0.1$
obtained in \cite{BKM99}
from an analysis of electromagnetic pion form factor.
\vspace*{1mm}

The above results have been obtained within the framework of NLCs,
employing a single non-locality parameter, viz., $\lambda_q^2$,
fixed at the so-called ``standard'' value of $0.4~\gev{2}$
\cite{BI82}. To examine the role of this parameter in determining
the structure of the model DAs, we constructed another set of
amplitudes, $\varphi_2(x)$, fixing it at the higher, yet
admissible, value $\lambda_q^2=0.5~\gev{2}$ \cite{Piv91}. Again,
an ``optimum'' DA, $\varphi_2^{\text{opt}}(x)$, was determined
with coefficients $a^{\text{opt2}}_2= +0.126 ,\,
a^{\text{opt2}}_4= -0.091$ with a profile similar to that of
$\varphi_1^{\text{opt}}(x)$, as Fig.~\ref{fig:DArange}(b) reveals.
Note, however, that now the envelope of the ``bunch'' becomes
quite close to the asymptotic DA, i.e., ``dromedary-like (convex)
at $x=1/2$ and correspondingly it exhibits less suppression in the
endpoint region $x=0,1$. We note in passing that a suppression of
the endpoint region as strong as possible (for a discussion we
refer to \cite{SSK99}) is important in order to improve the
self-consistency of perturbative QCD in convoluting the pion DA
with the specific hard-scattering amplitude of particular
exclusive processes. As before, the values of the inverse moment
$\va{x^{-1}}^{\rm SR}_{\pi 2}$, extracted from the new SR, and
those following from the model $\varphi_2(x)$, are mutually
consistent: \be 3.05 \leq \va{x^{-1}}_{\varphi_2} \leq 3.22
 ~~~~\text{vs}~~~~
   2.87 \leq \va{x^{-1}}^{\rm SR}_{\pi 2} \leq 3.51\ .
\ee 
It is worth remarking here that this ``camel-like'' structure of the
``best fit'' (alias, the optimum) DAs for both considered values of
the non-locality parameter $\lambda_{q}^{2}$, displayed in
Fig.~\ref{fig:DArange}(a,b) can actually be traced back to a
definite origin.
Crudely speaking, this shape structure is the net result of the
interplay between the perturbative contribution and the
non-perturbative term $\Delta\Phi_S(x)$ that dominates the rhs of the
SR in Eq.~(\ref{eq:SRDA}).
The fact that the function $\Delta\Phi_S(x)$ is not singular in $x$
and has a dip at the central point of the interval $[0,1]$ is also
reflected in the shapes of the DAs.

\section{QCD SR results vs CLEO data}
\begin{figure}[thb]
 \centerline{\includegraphics[width=\textwidth]{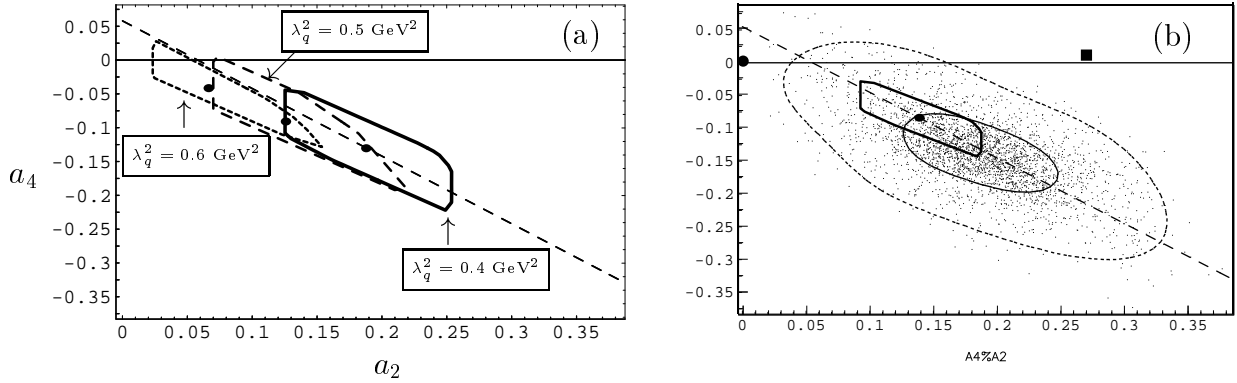}}
   \vspace*{1mm}\myfig{\label{fig:ogurec}}
    \caption{\footnotesize
   (a) The parameter space of ($a_{2}$, $a_{4}$) pairs (enclosed by a
   thick solid line), corresponding to the allowed values of the second
   and fourth Gegenbauer coefficients, calculated within the NLC-SR
   approach for three different values of
   $\lambda_{q}^{2}$ at $\mu=1~\Gev$.
   The optimum DA is denoted by a black dot.
   (b) shows how our estimate of the confidence region with
   $\lambda_{q}^{2}=0.4$~GeV${}^{2}$ at $\mu=2.4~\Gev$ overlaps
   with those displayed in Fig.~6 of Ref.\protect\cite{SchmYa99}.
   Bold dots in plot (b) mark the parameter pairs for the asymptotic
   (big black dot), Chernyak-Zhitnitsky (full square) and our
   optimum (with respect to the SR) DA (black dot).
   Contour lines show $68\%$ (solid line) and $95\%$ (dashed line)
   confidential regions, extracted by Schmedding and Yakovlev from
   $3000$ randomly chosen sets of CLEO data \protect\cite{CLEO98}.
   The small axis of the ``ellipses'' is generated by
   experimental-statistical uncertainties, whereas the large one
   results from theoretical-systematical uncertainties.}
\end{figure}

\textbf{Overlap of regions in} $\bbox{a_2, a_4}$ \textbf{plane.}
Let us now draw in Fig.~\ref{fig:ogurec}(a) the regions of
admissible pairs ($a_2$, $a_4$), determined in correspondence with
the error bars of the moments, in order to compare them with the
experimental constraints supplied by the recent high-precision
CLEO data \cite{CLEO98}. The latter are processed in the plot in
Fig.~\ref{fig:ogurec}(b) in the form of confidence regions,
extracted by Schmedding and Yakovlev \cite{SchmYa99} using a NLO
light-cone QCD SR analysis that (approximately) corresponds to an
average normalization point of $\mu=2.4$~GeV.


\vspace*{1mm}

The region, corresponding to the bunch of DAs, determined at the
``standard'' value $\lambda_q^2=0.4~\gev{2}$, and evolved to
$\mu=2.4$~GeV in Fig.~\ref{fig:DArange}(a), is enclosed in
Fig.~\ref{fig:ogurec}(a) by a slanted rectangular area.
Superimposing this area with the ``ellipses'' pertaining to a
$68\%$ (solid line) and a $95\%$ (dashed line) confidence region,
\cite{SchmYa99}, we show in Fig.~\ref{fig:ogurec}(b) to which
extent our estimates for the pion DAs agree with the CLEO data. In
particular, the central point
$(\bar{a}^{\text{opt}}_2,~\bar{a}^{\text{opt}}_4 )$ of the slanted
rectangle lies just on the $68\%$-region and the lower corner of it
``touches'' the central point $(a^{\ast}_2,~a^{\ast}_4 )$ of the
SY plot:
\ba \la{centre1}
 &a^{\ast}_2~=0.190,\quad
 &a^{\ast}_4~~~=-0.14 \ , \\
 \la{centre2}& \bar{a}^{\text{opt1}}_2\,=0.139,\qquad
& \bar{a}^{\text{opt1}}_4=-0.082 \ .
\ea 
Given the errors determined by SY, 
\be \la{errors} 
  a_2=a^{\ast}_2 \pm 0.04 (\text{stat}) 
                 \pm 0.09 (\text{syst}),
~~a_4=a^{\ast}_4 \pm 0.03(\text{stat}) 
                 \mp 0.09 (\text{syst}) \ , 
\ee 
one realizes that our theoretical estimates are quite compatible
with their experimental constraints. 
The region enclosed by the slanted rectangle 
bounded by the long-dashed line along the ``diagonal'' (see below) 
in Fig.~\ref{fig:ogurec}(a), corresponds to the bunch of DAs 
displayed in Fig.~\ref{fig:DArange}(b). 
Though still within the $95\%$ confidence region of SY, 
it is, however, mostly outside the central $68\%$ region. 
Finally, the third slanted rectangle 
limited by the short-dashed contour, and shifted along the ``diagonal'' 
to the upper left corner of the figure in Fig.~\ref{fig:ogurec}(a), 
corresponds to a {\it trial} NLC-DA with $\lambda_q^2=0.6~\gev{2}$. 
This value falls actually outside the standard QCD NLC-SR bounds 
in Eq.~(\ref{lambda}) for $\lambda_q^2$.
Remarkably, the image of this region in Fig.~\ref{fig:ogurec}(b)
(not displayed) would lie completely outside 
the central region as a whole. 
Therefore, we conclude that the CLEO data does not prefer 
the value $\lambda_q^2=0.6~\gev{2}$, 
in full agreement with previous QCD SR estimates.\footnote{%
\footnotesize{A very interesting detail in
Fig.~\ref{fig:ogurec}(a) is that the intersection set of all three
admissible regions represented by the slanted rectangles is
located around (and contains) the optimum point for the value
$\lambda_q^2=0.5~\gev{2}$, though its image in
Fig.~\ref{fig:ogurec}(b) lies outside, but close to the upper
boundary of the $68\%$ SY region.}}

\textbf{Role of the ``diagonal''.} By measuring the $\gamma +
\gamma^{\ast} \to \pi$ form factor, $F_{\gamma \gamma^{\ast}\pi
}$, one mainly probes the pion inverse moment $\va{x^{-1}}_\pi=3
(1+ a_2 + a_4 + \ldots)$. To illustrate this, let us apply the
leading twist NLA  expression \cite{KMR86} to the $Q^2 F_{\gamma
\gamma^{\ast}\pi }$, \be \la{radcorr} \Ds \frac{3}{4\pi}Q^2
\left(F_{\gamma \gamma^{\ast}\pi } \approx C\otimes \varphi_{\pi}
\right)
 = I
 = 3 \left[1 +  a_2 + a_4 + \ldots
 + \as( \Delta_0 + a_2\Delta_2 + a_4\Delta_4 + \ldots) \right] \ .
\ee 
Here $C$ is a coefficient function in the factorized amplitude 
for the process at NLO; 
$\Delta_n$ gives the size of the leading radiative corrections 
to the contribution of the Nth Gegenbauer eigenfunction 
entering the expansion of $\varphi_{\pi}(x)$. 
It should be noted that this standard factorization formula 
is not well established 
when one of the photons in the process is on-shell. 
Nevertheless, for demonstration purposes this approach is good enough. 
The correspondence 
\be \la{represent}
I(\mu^2)=\va{x^{-1}}_\pi(\mu^2)+3\as(\mu^2) \Delta_0 
\ee 
is slightly smeared by radiative corrections due to higher
eigenfunctions in Eq.~(\ref{radcorr}). 
Therefore, constraints 
imposed by $F_{\gamma \gamma^{\ast}\pi}$ measurements 
are in reality constraints for the combination $a_2 + a_4 + \ldots$,
modulo contributions from radiative corrections and higher twists.
Due to this reason, the admissible region for the desired points
$(a_2, a_4)$ in~\cite{SchmYa99}, (Fig.~\ref{fig:ogurec}(b)), 
is, so to speak, stretched along a ``diagonal'' 
defined by $a_2+a_4 = {\rm const}$. 
The value of this constant can be easily extracted
from the data with a reasonable accuracy,\footnote{\footnotesize{At
the same time, the theoretical-systematical errors to every pair of
values of $a_2, a_4$ appear to be very large, see
Eq.~(\ref{errors}).}}
using Eqs.~(\ref{centre2}), (\ref{errors}),
to obtain 
\be \la{diag-CLEO}
 a_2 + a_4 = 0.05 \pm 0.07
\ee 
(large systematic errors cancel!)
that should be compared with our value
\be \la{diag-SR}
 \bar{a}_2 + \bar{a}_4 = 0.056 \pm 0.033\ .
\ee 
Hence, we see that an ``experimental'' quantity
(Eq.~(\ref{diag-CLEO})) expressively agrees with those obtained in
the NLC-SR (cf. Eq.~(\ref{diag-SR})). 
On the other hand, 
the direct estimate for $\va{x^{-1}}_\pi$ from this SR 
(at $\mu^2 \sim 1~\gev{2}$ in Table 1) 
is also in agreement with both of these estimates, 
$\va{x^{-1}}_\pi /3-1\approx 0.1\pm 0.1$. 
The evolution of this quantity to the scale $\mu^2=(2.4~\Gev)^2$ 
should eventually make the agreement more pronounced. 
To complete these considerations, 
let us mention that the empirical SY estimate,
introduced in~\cite{SchmYa99} for a ``deformed diagonal'' in the
$(a_2, a_4)$-plane 
\be \la{ShYa-CLEO}
 a_2 + 0.6 a_4 = 0.11 \pm 0.03\ ,
\ee 
is also satisfied by the values of the central point of our admissible
region, namely,
\be
 \bar{a}_2 + 0.6 \bar{a}_4   = 0.09\pm 0.039\ .
\ee 

We emphasize that the radiative corrections in
Eqs.~(\ref{radcorr}), (\ref{represent}) are particularly important
because it turns out that they dominate over the ``non-perturbative''
$a_2,~a_4$ contribution.
To make this point more apparent, let us consider the asymptotic DA in
NLO approximation \cite{Mul94} (see Eqs.~(\ref{asympt}), (\ref{invers})
in Appendix B).
Then, we find that this eigenfunction supplies the main contribution to
$I$ and $\as \Delta_0$:
\be \la{radcorr0}
\Ds \as \Delta_0
    = \frac{\as(\mu^2)}{4\pi}\left[-5C_F
    + \left(2C_F - \frac{ b_0}{3}\right)\right]
    = - 7~\frac{\as(\mu^2)}{4\pi}\approx -0.17\ .
\ee 
The first contribution, $-5C_F$, originates from the coefficient
function (see \cite{MuR97}), while the second one,
$\left( 2C_F - b_0/3 \right)$, from the modification of the
asymptotic DA at NLO \cite{Mul94}.
The final estimate, $-0.17$, corresponds to the value
$\as/(4\pi)=0.024$ \cite{LD92} at the normalization point
$\mu = 2.4~\Gev$.
We see that the estimate for $\as \Delta_0$ is indeed 3 times larger
than those for $a_2+a_4$, considered above, and for that reason it
controls the level of accuracy in the determination of the
latter.\footnote{%
\footnotesize{An analysis in leading approximation, performed
in~\cite{RR96} for the phenomenological value $I \approx 2.4$,
leads to the conclusion that $I\approx \va{x^{-1}}_\pi$, so that
the associated DA is forced to be narrower than the asymptotic
one. But in NLO this estimate transforms into
Eqs.~(\ref{radcorr}), (\ref{represent}) and therefore one can
conclude that the value of $\va{x^{-1}}_\pi$ is larger than 3,
with the pion DA being broader than the asymptotic one. Indeed,
Fig.~\ref{fig:ogurec}(b) demonstrates exactly this fact in detail.
In the same context we mention that a narrower DA than the
asymptotic one was also used in the calculation of $F_{\pi\gamma}$
in \cite{SSK00} yielding a prediction somewhat below the CLEO data
and leading to the suggestion that the shape of the ``true'' pion
DA must be broader than that of the asymptotic one.}} The twist-4
contribution to the form factor is also important (see
\cite{SchmYa99}), being of the same order as $a_2+a_4$ but having
the opposite sign. The overall radiative correction to the
contribution of still higher eigenfunctions is also negative and
small in comparison to the estimate in Eq.~(\ref{radcorr0}),
$$\frac{\as(\mu^2)}{4\pi}\cdot (a_2\Delta_2 + a_4\Delta_4)
  \approx 0.024 \cdot (-0.58) = -0.014\ .$$
To further improve the theoretical accuracy of extracting $a_2$
and $a_4$ (from experimental data), one should first obtain an
estimate for $C\otimes \varphi_{\rm as}$ in next to NLO
approximation because this contribution will certainly dominate in
that order.

\section{Comparison with results from different sources.}
In this section we compare the results of our analysis with new
evidences for the form of the pion DA, following from hadron
phenomenology, lattice simulations and instanton-based models.
However, keep in mind that the information provided from these sources
bears uncertainties the size of which is difficult to estimate.

\begin{figure}[thb]
 \centerline{\includegraphics[width=\textwidth]{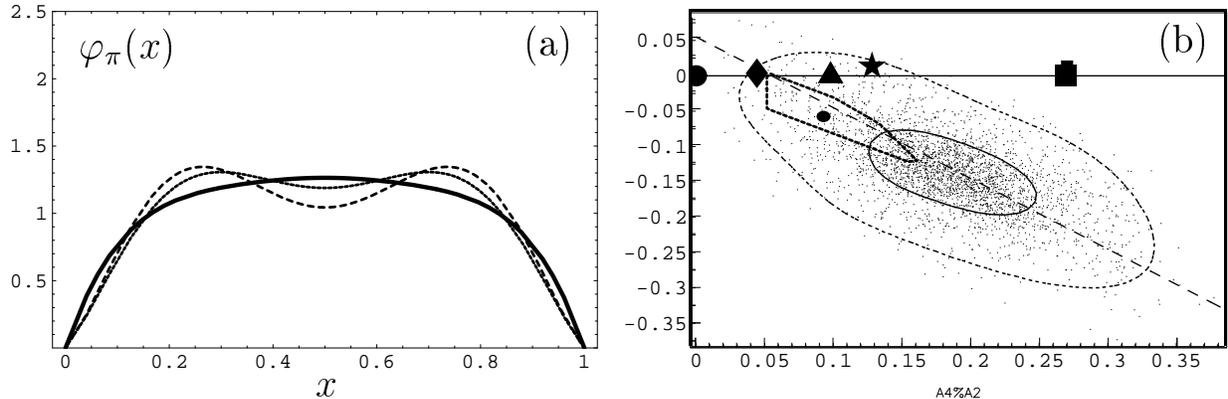}}
   \vspace*{1mm}\myfig{\label{fig:summary}}
    \caption{\footnotesize
   (a) Selected profiles of $\varphi_{\pi}(x,\mu^2=8~\gev{2})$.
   The short (long) dashed line corresponds to our ``optimum'' DA
   model,    $\varphi_1^{\text{opt}}(x)$ at $\lambda_q^2=0.4~\gev{2}$
   ($\varphi_2^{\text{opt}}(x)$ at $0.5~\gev{2}$) and the solid line
   to the fit to the model deduced from the experimental data on di-jet
   production in pion-nucleon interactions~\protect\cite{Ash00}.
   (b) This plot is obtained by inserting into Fig.~6 of
   Ref.\protect\cite{SchmYa99} the confidence region determined for the
   $\varphi_2$-bunch (at $\lambda_q^2=0.5~\gev{2}$) and confined
   in the slanted rectangle.
   The E791 Collaboration fit\protect\cite{Ash00},
   Eq.~(\protect\ref{E791}), is denoted by a black star, whereas
   the Dalley model, Eq.~(\protect\ref{Dalley}), is marked by a black
   triangle.
   All model DAs shown are evolved to $\mu=2.4~\Gev$.}
 \end{figure}

\textbf{DA from} $\bbox{\pi^{-}}$ \textbf{into di-jets via diffractive
dissociation.}
It is tempting to compare the results obtained and discussed above
with the intriguing interpretation of experimental data on di-jet
production in pion-nucleon interactions \cite{Ash00}.
This process was suggested already in~\cite{FMS93} as a tool to measure
the shape of the pion DA directly from the data via diffractive
dissociation into two jets.
Such an experiment was carried out very recently by the Fermilab E791
Collaboration \cite{Ash00} by measuring the transverse momentum
distribution of the diffractive di-jets in two windows of $k_t$:
$1.25 \leq k_t \leq 1.5$~\Gev and $1.5 \leq k_t \leq 2.5$~\Gev.
We have found that their experimental deduction for the shape of the
pion DA in the lower window can be well approximated by a model with
two Gegenbauer coefficients to read
\be
\la{E791}
 \varphi_{\pi}^{\text{fit}}(x,Q^2) = \varphi^{\rm as}(x)
 \left[1+  a^{\text{fit}}_2(Q^2) \cdot C^{3/2}_2(2x-1)
        +  a^{\text{fit}}_4(Q^2) \cdot C^{3/2}_4(2x-1) \right]\ ,
\ee 
with $\chi^2 \approx 10^{-2}$ and
$a^{\text{fit}}_2=0.121,~a^{\text{fit}}_4=0.012$ at $Q^2 \approx 8
~\gev{2}$. Despite the large ``diagonal'' value
$a^{\text{fit}}_2(\mu^2) + a^{\text{fit}}_4(\mu^2) \approx 0.143$
(at $\mu=2.4~\Gev$), resulting from this  model, it is still
consistent with the SY constraints, as one realizes by comparing
it with the values given in Eqs.~(\ref{diag-CLEO}),
(\ref{diag-SR}). The symbol ``star'', associated with
$\varphi_{\pi}^{\text{fit}}(\mu^2)$, is located in
Fig.~\ref{fig:summary}(b) on the upper boundary of the SY $95\%$
region and above the diagonal. Fig.~\ref{fig:summary}(a) shows the
profile of the E791-fit (thick solid line) in comparison with our
``optimum'' DAs for $\lambda_q^2=0.4$ (dashed line) and
$\lambda_q^2=0.5~\gev{2}$ (solid line), both evolved to
$Q^2\approx 8~\gev{2}$. Taking into account that the accuracy of
the data is not better then $10\%$, one is tempted to conclude
that the DA shape from the di-jet data is in fact not far away
from the $\varphi_2$-bunch. Note that the authors of the second
paper in \cite{FMS93} have discussed this treatment of di-jet
production data and concluded that the interpretation of them with
respect to the pion DA is yet questionable. Indeed, inspection of
their Fig.~3 \cite{Ash00} reveals that the results (data) for
$x<~0.1$ and $x>~0.85-0.9$ are different, exhibiting the
limitations of the accuracy of their fit in the endpoint regions.

\textbf{Transverse lattice.}
Quite recently, Dalley \cite{Dal01} has used the data points in a
transverse-lattice calculation to deduce the shape of the pion DA and
approximated it by the following analytic form
\be \la{Dalley}
 \varphi^{\text{lat}}(x,\mu^2_0) = \varphi^{\rm as}(x)
 \left[1+ 0.133 \cdot C^{3/2}_2(2x-1)
        \right]
\ee 
at $\mu^2_0 \sim 1~\gev{2}$. The shape parameters $a_2, a_4$ of
this DA are inside the admissible $95\%$ region in
Fig.~\ref{fig:summary}(b) (black triangle) and somewhat closer to
the diagonal than the fit of E791. Because of several
approximations in this calculation, the quality of the deduced fit
is unknown and as a result the errors in the coefficients are
presumably large. Hence, we may regard this model DA as being only
qualitatively consistent with the $\varphi_2$-bunch.

Two things are worth to be remarked. First, this new model fit is
in conflict with the DA shape predicted in~\cite{Burk99} and more
recently in \cite{BuSe01} using the same Hamiltonian and Fock
space truncation, but employing continuous basis functions rather
than the discrete light-cone quantization applied by Dalley.
Second, if the shape obtained is very close to the asymptotic one,
then it is again ruled out by the CLEO data, as it is outside the SY
confidence regions. For the sake of completeness, we mention in
this context that other lattice attempts \cite{Lat91,Lat00} yield
unrealistically small values of the moment $\va{\xi^2}_{\pi}$.

\vspace*{1mm}

\textbf{Instanton-induced models.} Instanton-induced models,
worked out in \cite{PPRWG99,ADT00}, usually lead to DA shapes
close to the asymptotic one. As a consequence, such DAs are
definitely ruled out by the SY constraints shown in
Fig.~\ref{fig:summary}(b). However, these DAs satisfy ``a softer
constraint'', also derived by SY in \cite{SchmYa99} (see Fig.~7
there). In fact, the confidence regions derived this way, are
somewhat enlarged with the effect that the asymptotic DA is just
``inside'' the larger region and positioned at the upper end of
the diagonal. Note that instanton-induced models prefer a higher
value of the vacuum quark virtuality: $\lambda_q^2 \geq
0.6~\gev{2}$. For such a value of the parameter $\lambda_q^2$
(unrealistically large for QCD SRs), the condensate contribution
on the rhs of Eq.~(\ref{eq:SRDA}) becomes small in comparison to
the perturbative one. As a consequence of this, the corresponding
DA appears to be closer to the asymptotic one, as we have already
indicated (cf. Fig.~\ref{fig:ogurec}(a), where one sees that the
center of the slanted rectangle for this $\lambda_q^2$ value is
much closer to the $a_4$ axis and hence to the asymptotic DA.)

A separate case of an instanton-induced pion DA, consistent with
the SY constraints, has been derived by Praszalowicz
\cite{MPra01}. We included this model in Fig.~\ref{fig:summary}(b)
by the label ``full diamond''.

\section{Conclusions}

1. We performed an accurate analysis of QCD SR with NLCs for the
pion DA and obtained admissible sets of them with respect to the
SR constraints for different values of the non-locality parameter
$\lambda_q^2$. Confronting these determined DA spectra with the
constraints obtained by SY in \cite{SchmYa99} from the
high-precision CLEO experimental data \cite{CLEO98}, we showed
that they satisfy them {\it quantitatively} (cf.
Fig.~\ref{fig:ogurec} and Fig.~\ref{fig:summary}(b)).

2. We found that the predictions for the profile of the pion DA
deduced from the di-jet production (E791 Collaboration)
\cite{Ash00} and those extracted from the simulation  on a
transverse lattice \cite{Dal01} are consistent only with the 95\%
SY constraints ($2\sigma$-deviation criterium), but not with the
68\% SY constraints ($1\sigma$-deviation criterium). Moreover,
they are located near the boundary of the $2\sigma$-region
obtained with the standard error estimation. As shown in
Fig.~\ref{fig:summary}(b), the di-jet prediction is not too far
from the $\varphi_2$-bunch region (with $\lambda_q^2 =
0.5~\gev{2}$), though a quantitative comparison calls for a higher
accuracy of the data processing. On the other hand, the lattice
result qualitatively also agrees with the set of our DAs (the
$\varphi_2$-bunch) (see Fig.~\ref{fig:summary}(b)), but a more
detailed comparison is obscured by the unknown quality of this
method.

3. The existing predictions from instanton-induced models are too
close to the asymptotic DA and therefore they are only consistent
with the ``softened'' version of the SY constraints (see for
details in \cite{SchmYa99}), except for the new model by
Praszalowicz \cite{MPra01}, which is just outside the confidence
region of the $\varphi_2$-bunch and on the boundary of the
$95\%$-region of SY. \vspace*{3mm}

\noindent \textbf{Acknowledgments.} This work was supported in
part by the Russian Foundation for Fundamental Research
(contract 00-02-16696), INTAS-CALL 2000 N 587, 
the Heisenberg--Landau Program (grants 2000-15 and 2001-11),
and the COSY Forschungsprojekt J\"ulich/Bochum.
We are grateful for discussions to A.\ Dorokhov, L.\ Frankfurt,
P.\ Pobylitsa, V.\ Petrov, M.\ Polyakov, M.\ Praszalowicz, and
M.\ Strikman.
Two of us (A.B. and S.M.) are indebted to Prof.\ Klaus Goeke
for the warm hospitality at Bochum University, where this
work started.

\begin{appendix}
\vspace*{12mm}
\appendix
\vspace{-3mm}

\section{Expressions for nonlocal contributions to SR}
 \renewcommand{\theequation}{\thesection.\arabic{equation}}
  \la{subs-A.1}\setcounter{equation}{0}

The form of contributions of NLCs to the OPE on the rhs of
Eq.~(\ref{eq:SRDA}) depends on the modelling of NLCs. At the same
time, the final results of the evaluation of the SRs show
stability against variations of this modelling, provided the scale
of the average vacuum quark virtuality $\lambda_q^2$ is fixed.
Here, we used the model (delta-ansatz) suggested in
\cite{MR89,BR91} and used extensively in \cite{BM98,BM00}. This
model leads to a Gaussian decay of the scalar quark condensate
$M_S(z^2)$. For the four-quark condensate, the factorization
ansatz is applied to reduce its contribution to a pair of scalar
condensates. In the NLC approach this may lead to an overestimate
of the four-quark condensate contribution,
$\Delta\Phi_S\left(x;M^2\right)$, because it evidently neglects
the correlation between these pairs. Below, $\Delta \equiv
\lambda_q^2/(2M^2)$, $\bar\Delta\equiv 1-\Delta$:
\ba \label{qs}
 \Delta\Phi_S\left(x;M^2\right)
  &=&\frac{A_S}{M^4}
      \frac{18}{\bar\Delta\Delta^2}
       \Bigl\{
        \theta\left(\bar x>\Delta>x\right)
         \bar x\left[x+(\Delta-x)\ln\left(\bar x\right)\right]
       +  \xx + \nonumber \\
&&\qquad\qquad  +
\theta(1>\Delta)\theta\left(\Delta>x>\bar\Delta\right)
         \left[\bar\Delta
              +\left(\Delta-2\bar xx\right)\ln(\Delta)\right]
         \Bigr\}\ .
 \la{eq:phi_s} 
\ea 
Instead of (A.5) in \cite{BM98}, we have
\ba
 \Delta\Phi_{T_1}\left(x;M^2\right)
  &=& -\frac{3 A_S}{M^4}\theta(1>2\Delta)\left\{
       \left[\delta(x-2\Delta) - \delta(x-\Delta)\right]
       \left(\frac1{\Delta} - 2\right)
     + \theta(2\Delta>x) \cdot\nn \right.\\ &&\left.
    \theta(x>\Delta)
  \frac{\bar x}{\bar\Delta}
  \left[\frac{x-2\Delta}{\Delta \bar\Delta} \right]
  \right\} + \xx\ .
\ea 
Here, $\Ds A_S=\left(8\pi/81\right)\langle\sqrt{\as}\bar
q(0)q(0)\rangle^2$, whereas for the quark and gluon condensates we
use the standard estimates $\langle\sqrt{\as}\bar
q(0)q(0)\rangle\approx (-0.238~\Gev)^{3}$, $\Ds \langle\as
GG\rangle/12\pi\approx 0.001~\gev{4}$ \cite{SVZ}.

\section{ Radiative corrections }
 \la{subs-A.2}
  \setcounter{equation}{0}

The radiative corrections to the correlator and to the pion DA are
\ba \la{eq:rcpi}
 \rho^{\rm pert}(x,s)
  &=& 3 x\bar x
     \left\{1+ a_s
      \left[C_F\left(2 - \frac{\pi^2}{3} + \ln^2(\bar x/x)\right)
+ C_F\left(2 \ln\left(s/\mu^2\right)+3\right)\right]\right\}
  \frac{1}{2\pi^2}\ ,
\\
 \la{asympt}
\varphi_{\rm as}^{\rm NLA}(x) &=&  6x\bar x
    \left\{1+ a_s
    \left[C_F \left(2 - \frac{\pi^2}3 +
    \ln^2(\bar x/x)\right) 
  + b_0\,\left(\ln(x\bar x) + \frac5{3}\right)\right]\right\}\ , \\
\la{invers} \va{x^{-1}}_{\rm as} &=& 3 \left[1 + a_s \left(2C_F -
\frac1{3}b_0\right)\right]\ . 
\ea 
Here, $b_0$ is the first $\beta$-function coefficient, $b_0=(11/3)
C_A - (4/3) Tr N_f$; $ a_s=\alpha_s(\mu^2)/4\pi$, $\Lambda^{\rm
NLO}_{(3)}\approx 0.465~\Gev$, $\alpha_s^{\rm
NLO}(m_\tau^2)\approx 0.358 $, see, e.g., \cite{LD92}. We perform
a 2-loop evolution of the pion DA following the approach presented
in the first paper in \cite{KMR86} and using the so-called
``optimized MS-scheme'' \cite{BRS00}.
\end{appendix}


\newpage
\centerline{\bf Erratum}
\vspace*{3mm}

We used a wrong formula, Eq.\ (B.1),
for the radiative corrections to the spectral density 
of the axial-axial correlator. 
The correct one should read as follows ($\bar{x}\equiv 1-x$):
\begin{eqnarray} 
 \rho^{\rm pert}(x,s)
  &=& 3\, x\bar x
     \left[1+ a_s\, C_F\left(5 - \frac{\pi^2}{3} + \ln^2(\bar x/x)\right)
     \right]
  \frac{1}{2\pi^2}\,.
\end{eqnarray}
We reanalyzed our sum rules using the correct expression
and obtained new parameters for
the phenomenological spectral density in the axial channel:
\begin{eqnarray} 
  \rho^\text{phys}(s)
  &=& f_{\pi}^2\delta(s)
   +  f_{A_1}^2\delta(s-m_{A_1}^2)\,,
\end{eqnarray}
where $A_1$ in fact represents a mixture 
of the real $A_1$ and $\pi'$ mesons.
The optimum stability of the analyzed sum rules 
is achieved for $s_{0}=2.25$~GeV${}^{2}$
and $m_{A_1}^2=1.616$~GeV${}^{2}$.
The error induces mainly a shift 
in the determination of the decay constants
of the pion and its excitations;
it does practically not influence the values of the extracted moments 
of the pion distribution amplitude,
as one can see from Table~1,
where we collect the values of these quantities in comparison 
with the published ones. 
\begin{table}[h]
\footnotesize
\begin{tabular}{|c|c|c|c|c|c|c|}\hline
 $f_{\pi}\left(\text{GeV}\right)$
   &$\hspace{0.1mm}N=2\hspace{0.1mm}$
     &$\hspace{0.1mm}N=4\hspace{0.1mm}$
       &$\hspace{0.1mm}N=6\hspace{0.1mm}$
         &$\hspace{0.1mm}N=8\hspace{0.1mm}$
           &$\hspace{0.1mm}N=10\hspace{0.1mm}$
             &$\hspace{0.1mm}\va{x^{-1}}\hspace{0.1mm}$\\ \hline
 ${\strut\vphantom{\vbox to 6mm{}}}$
 $\hspace{1mm}0.131(8)\hspace{1mm}$
   & $\hspace{1mm}0.265(20)\hspace{1mm}$
     & $\hspace{1mm}0.115(12)\hspace{1mm}$
       & $\hspace{1mm}0.061(8)\hspace{1mm}$
         & $\hspace{1mm}0.037(5)\hspace{1mm}$
           & $\hspace{1mm}0.024(4)\hspace{1mm}$
             & $\hspace{1mm}3.35(32)^\text{SR}\hspace{1mm}$\\ \hline
 ${\strut\vphantom{\vbox to 6mm{}}}$
 $0.137(8)$
   & $0.266(20)$
     & $0.115(11)$
       & $0.060(7)$
         & $0.036(5)$
           & $0.025(4)$
             & $3.35(30)^\text{SR}$\\ \hline
\end{tabular}
\vspace*{3mm}

\noindent\textbf{Table 1}. The moments 
$\langle \xi^N \rangle_{\pi}(\mu^2)$ determined at $\mu^2 \sim 1.35~\gev{2}$ with
associated errors put in parentheses.
The old values are given in the first row
and the new ones in the second row.
\end{table}
\vspace*{3mm}

Table~2 represents the changes for the effective $A_1$-meson
decay constants and moments:
\begin{table}[h]
\footnotesize
\begin{tabular}{|c|c|c|c|c|c|c|}\hline
 $f_{A_1}\left(\text{GeV}\right)$
   &$\hspace{0.1mm}N=2\hspace{0.1mm}$
     &$\hspace{0.1mm}N=4\hspace{0.1mm}$
       &$\hspace{0.1mm}N=6\hspace{0.1mm}$
         &$\hspace{0.1mm}N=8\hspace{0.1mm}$
           &$\hspace{0.1mm}N=10\hspace{0.1mm}$
             &$\hspace{0.1mm}\va{x^{-1}}\hspace{0.1mm}$\\ \hline
 ${\strut\vphantom{\vbox to 6mm{}}}$
 $\hspace{1mm}0.210(17)\hspace{1mm}$ 
   & $\hspace{1mm}0.21(2)\hspace{1mm}$
     & $\hspace{1mm}0.116(12)\hspace{1mm}$
       & $\hspace{1mm}0.078(8)\hspace{1mm}$
         & $\hspace{1mm}0.055(6)\hspace{1mm}$
           & $\hspace{1mm}0.042(5)\hspace{1mm}$
             & $\hspace{1mm}3.6(4)^\text{SR}\hspace{1mm}$\\ \hline
 ${\strut\vphantom{\vbox to 6mm{}}}$
 $0.221(20)$
   & $0.21(2)$
     & $0.113(12)$
       & $0.076(8)$
         & $0.055(6)$
           & $0.040(5)$
             & $3.5(4)^\text{SR}$\\ \hline
\end{tabular}
\vspace*{3mm}

\noindent\textbf{Table 2}. The moments $\langle \xi^N \rangle_{A_1}(\mu^2)$ 
determined at $\mu^2 \sim 1.35~\gev{2}$.
\end{table}
\vspace*{3mm}

   The old and new parameter regions of ($a_{2}$, $a_{4}$) pairs, 
corresponding to the allowed values 
of the second and fourth Gegenbauer coefficients, 
calculated with the corrected sum rules in comparison with the published ones
for the value $\lambda_{q}^{2}=0.4$~GeV$^2$ at $\mu^2\sim 1.35~\gev{2}$ 
are given in Fig.~1.
\vspace*{1mm}
 \begin{figure}[thb]
 \centerline{\includegraphics[width=0.5\textwidth]{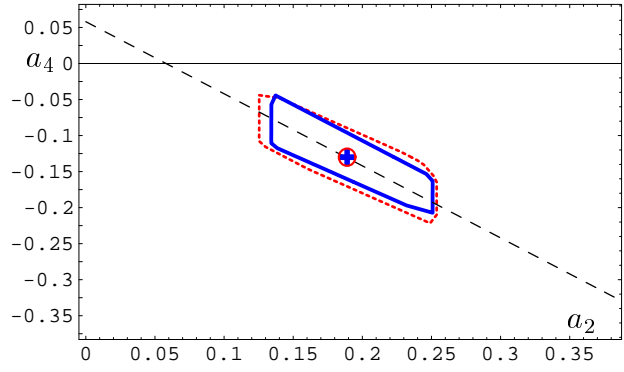}}
   \caption{\footnotesize
   Corrected ($a_{2}$, $a_{4}$) area (blue solid line) in comparison
   with the published one (red dotted line).
   The blue cross marks the new  position of the BMS distribution amplitude,
   while the red circle denotes the old one.}
\end{figure}

Given that the two ($a_{2}$, $a_{4}$) areas almost coincide 
and the BMS distribution amplitude is only infinitesimally shifted
none of the results of our analysis
is affected.

\end{document}